# HOW TO (DIS-)ASSEMBLE A PLANETARY SYSTEM
## by turning a video game into an educational game

S. Iovenitti [*1,2] and L. Perri [2]

**Abstract.** Nowadays, computer simulations have reached amazing graphic results, providing awesome visual descriptions explaining the evolution of very complex physics phenomena, implementing the laws of Nature. Sometimes, without a scientific guide, it is difficult to distinguish between simulation and reality, but the usage of advanced and spectacular animations is a powerful way to convey both accurate scientific information and passion for science. According to this, we decided to develop an innovative scientific workshop based on an unconventional instrument: a video game. We used a commercial software called Universe Sandbox, a physics-based space simulator where astrophysics objects can be created, manipulated and followed in their evolution. We prepared a set of dynamic scenes to take the public through a virtual space journey whose main topics were planetary formation, orbits, habitable zones, types of stars, stellar explosions, and asteroid collisions. The interaction with the audience was at the heart of the experience: every question, suggestion, or idea was inserted in the simulation in real-time. This workshop was first proposed to the national festival of science in the city of Bergamo, in Italy, and we report our experience in this contribution. This activity is very easy to be replicated everywhere, both online and in-presence.

### 1. Introduction

It is well known that video games can play a positive role in increasing the awareness of important issues, provided that the game experience is followed by structured discussion, supervised by an expert guide [1]. This idea led us to develop a new concept, expanding the idea of a gaming session into an exploration involving multiple players, under the guidance of a scientist, across possible scenarios into a virtual space environment. This is the foundation of *"How to dis-assemble a planetary system"* [2], an innovative interactive scientific workshop based on the videogame Universe Sandbox [3]: a commercial software able to reproduce different astronomical systems and phenomena, implementing physics laws with spectacular graphic animations. Actually, in the workshop we do not play the video game with any specific goal: the attention is focused on modifying the universe (as it is expressed in the title), in order to discover its behaviour and its laws. Overall, as the amazing graphic environment constitutes a powerful way to convey passion for science together with accurate astrophysics knowledge [4], this workshop can be an innovative and effective strategy to communicate astronomy to a large public.

### 2. A videogame for communicating astronomy?

Universe Sandbox is a physics-based space simulator, a commercial software that can be bought on the most common video game platforms (e.g. Steam [5]). It merges gravity, thermodynamics and other physics laws in a virtual environment that can be created from scratch and populated. Under the guidance of an astrophysicist, this software can be used not only for entertainment, but also to spotlight lots of peculiar phenomena that are embedded in the simulations. For this reason, we believe that such a video game can be an effective tool for communicating astronomy, presenting the following several advantages. (i) The graphic environment allows us to show the true scale-relations of astronomy, that often people can't conceive simply because they are completely out of our range of experience and imagination. (ii) The usage of a video game avoids the feeling "it's too difficult for me" that lots of people, mostly young students, still perceive regarding astronomy and astrophysics. (iii) A videogame is a new and unconventional tool to be exploited for communication purposes and hence this engaging activity is "appealing" to the public.

On the other hand, there are also some critical issues. The software is very demanding from the point of view of the computational resources, hence a very high-performance machine is mandatory: if the minimum system requirements are not matched by the

1 Università degli Studi di Milano – Physics department.
   Via G. Celoria 16, 20133 Milano (MI), Italy.
2 INAF – Osservatorio Astronomico di Brera
   Via E. Bianchi 46, 23807 Merate (LC), Italy.
* Corresponding author: simone.iovenitti@inaf.it

computer, the graphics will not be fluid. Furthermore, since simulations are very complex and based on a sequence of approximations, the behavior of the software can be different even if input parameters are the same: when the audience wants to "see again" a simulation it may be impossible.

### 3. Structure of the workshop

We designed our activity in order to set the interaction with the audience at the heart of the experience. In the virtual environment of Universe Sandbox it is possible to create different astrophysics systems and consider their evolution in time or change their parameters. The software is operated by the scientific guide, performing the simulations, but it is the audience to decide what to do and why. It all started with a set of dynamic scenes, prepared in advance, whose main topics were planetary formation, orbits, habitable zones, types of stars, stellar explosions, and asteroid collisions. After this introduction, the audience is involved in creating new scenarios: every suggestion is inserted in real-time into the environment of Universe Sandbox by the guide. Similarly, the questions were not only answered by words, but with a simulation on purpose, created live! Even the fanciest and odd ideas were accepted, exploring the effects of physics laws in unlikely contests: collisions between planets, comets interactions, planetary systems around black holes, frozen or burning worlds, and many other strange situations. Each time the simulations were analyzed under a scientific perspective and connections with well-known real cases were discussed in detail in order to understand deeply the underlying physics process.

### 4. Our experience

Our activity was first proposed in 2020 at the national festival of science in the city of Bergamo, in Italy. That area was heavily affected by the COVID-19 pandemic, so our activity was delivered completely online, using a remote meeting connection and sharing the screen of the computer that was running the software. Both school groups (primary to high school) and the general public participated in this activity, and we had excellent feedback from all of them. Several schools chose to buy the software themselves and asked us to replicate the workshop for their future students. Thanks to the help of the Planetarium of Milan, we extended the duration of the activity for the first five months of 2021 and we planned several replicas until the end of the year.

We verified that the audience learned a lot of important notions about astronomy in our workshop, but the most important take-home message has been another: *we can use astronomy*. In fact, our activity demonstrated that astronomy is not only a descriptive science, because we showed astronomy in action: by knowing scientific laws, we can understand new phenomena, make predictions, simulate processes, and create new possible scenarios that may or may not exist elsewhere. In this context, to play with astronomy is not only a creative process but could be intended as an experimental activity, almost a hands-on experience. Another remarkable aspect of our experience is that this activity puts astronomy in connection with earth sciences, biology, and climate. In particular, we put in evidence that slight changes can destroy the conditions of a whole planet, leading people to be aware of the fragility and uniqueness of the Solar System: this feeling constitutes an immediate answer to the "what is it for?" problem in communicating astronomy.

### 5. Conclusion

With Universe Sandbox it is possible to create a workshop that does not require any complicated setup or large expense (the license of the software is very cheap). This activity can be easily replicated everywhere, in every context, and it can be delivered both online and as an in-presence activity, using a wall projector, or multiple displays, or again a network of VR headset. We verified that it constitutes an engaging experience and an effective tool to convey accurate scientific knowledge together with entertainment and passion for science: a powerful mix for communicating astronomy.

### References

[1] Burnet, F. *"More Scientists and Less Surrogates."* Journal of Science Communication, vol. 9, no. 2, Jun. 2010, doi:10.22323/2.09020304
[2] youtu.be/t8FsglAhQrE
[3] universesandbox.com
[4] Dudo, Anthony, et al. *"Portrayals of Technoscience in Video Games: A Potential Avenue for Informal Science Learning."* Science Communication, vol. 36, no. 2, Apr. 2014, pp. 219–247, doi:10.1177/1075547013520240.
[5] store.steampowered.com